\documentclass[12pt,preprint]{aastex}
\shorttitle{Formation of eclipsing binary pulsars}
\shortauthors{Chen et al.}
\usepackage{natbib}

\begin{document}

\title{Formation of black~widows and redbacks
-- two distinct populations of eclipsing binary millisecond pulsars}

\author{Hai-Liang Chen\altaffilmark{1,2,3}, Xuefei Chen\altaffilmark{1,2}, 
Thomas M. Tauris\altaffilmark{4,5} and Zhanwen Han\altaffilmark{1,2}}
\altaffiltext{1}{Yunnan Observatories, 
the Chinese Academy of Sciences, Kunming 650011, China; chenhl@ynao.ac.cn} 
\altaffiltext{2}{Key Laboratory for the Structure and Evolution of Celestial Objects, the Chinese Academy of Sciences,
Kunming 650011, China} 
\altaffiltext{3}{University of the Chinese Academy of Sciences, Beijing 100049, China}
\altaffiltext{4}{Argelander-Institut f\"ur Astronomie, Universit\"at Bonn, Germany}
\altaffiltext{5}{Max-Planck Institut f\"ur Radioastronomie, Bonn, Germany}

\begin{abstract}

Eclipsing binary millisecond pulsars (the so-called black~widows and redbacks) 
can provide important information about accretion history, pulsar irradiation of their companion stars
and the evolutionary link between accreting X-ray pulsars and isolated millisecond pulsars.
However, the formation of such systems is not well understood, nor the difference in progenitor evolution between the 
two populations of black~widows and redbacks. Whereas both populations have orbital periods between $0.1-1.0\;{\rm days}$
their companion masses differ by an order of magnitude. 
In this paper, we investigate the formation of these systems via evolution of converging low-mass X-ray binaries
by employing the MESA stellar evolution code.
Our results confirm that one can explain the formation of most of these eclipsing binary millisecond pulsars using this scenario.
More notably, we find that the determining factor for producing either black~widows or redbacks is the efficiency
of the irradiation process, such that the redbacks absorb a larger fraction of the emitted spin-down energy of the radio pulsar
(resulting in more efficient mass loss via evaporation) compared to that of the black~widow systems.
We argue that geometric effects (beaming) is responsible for the strong bimodality of these two populations. Finally,  
we conclude that redback systems do not evolve into black~widow systems with time.
   
\end{abstract}


\section{Introduction}
It is widely accepted that millisecond pulsars (MSPs) are produced via the recycling scenario \citep{acrs82,bv91}. 
In this scenario, the MSPs are formed from the evolution of low-mass X-ray binaries (LMXBs) and intermediate-mass X-ray binaries
in which a neutron star (NS) accretes material and angular momentum from its companion star and thereby is spun up to become 
an MSP. During this process, the magnetic field of the NS decreases to $B\simeq 10^{7}-10^{9}\;{\rm G}$. 
The exact mechanism for this accretion-induced magnetic field decay is still unknown \citep[e.g.][]{sbmt90,kb97,zhan98,czb01}. 
In most of the cases, a binary system with an MSP and a helium white dwarf is left behind when the LMXB mass transfer ends. 
It was therefore a bit of a surprise when the first black~widow system was discovered, revealing that the MSPs 
in these black~widow systems have very low-mass ($0.02-0.05\;M_{\odot}$) companions \citep{fst88,sbl+96}.
Since the pulsars in these systems suffer from eclipses of their radio signals, it was immediately clear that the companions are 
semi- or non-degenerate stars which suffer from irradiation-driven mass loss \citep{krst88,rst89,rste89}.

\citet{kdb03,kbr+05} proposed that these eclipsing MSP systems are produced in globular clusters and subsequently ejected into the Galactic field.
In their scenario, the formation of eclipsing binary pulsars should go through two phases: in the first phase, a binary system consisting of an  
MSP and a helium white dwarf is formed via the evolution of an LMXB in a globular cluster; In the second phase, the helium white dwarf is replaced 
by a main sequence star, via an exchange encounter event, leading to ablation and matter expelled from the system caused by strong irradiation
of the new companion by the energetic MSP. 
As regard to the low number of eclipsing binary MSPs known  
in the Galactic field at that time, they suggested that eclipsing binary pulsars are mainly generated in the globular clusters and 
subsequently ejected into the Galactic field or the systems entered the field population in case the cluster itself was disrupted.

However, in the last few years a rapidly increasing number of MSPs have been found in the Galactic field.
Most of these discoveries are binary eclipsing MSPs in tight orbits ($P_{\rm orb} \la 24\;{\rm hr}$), the so-called spiders \citep{rob13,pmd+12,rfs+12,kbv+13,bvr+13}. 
A large fraction of these new MSPs are associated with \textit{Fermi} $\gamma$-ray sources \citep[e.g.][]{aaa+09,pgf+12}. 
As \citet{rob13} suggested, the population of eclipsing binary MSPs  
can be divided into two different classes: black~widows and redbacks. The black~widows have very low-mass companions ($M_2 \ll 0.1 M_{\odot}$),  
while the redbacks have companion masses of a few tenths of a solar mass ($M_2 \simeq 0.1-0.4\;M_{\odot}$). 
In view of the many recent discoveries, it seems that the model suggested by King~et~al. may have difficulties in explaining 
the large number of such binary systems in the Galactic field. 

Detailed stellar evolution modelling has previously demonstrated that black~widows form as the outcome 
of converging LMXB evolution \citep[e.g.][]{ef92,prp02,bdh12}. 
In this paper, we perform binary evolution calculations to study the formation of both redbacks and black~widows, 
and also investigate whether or not redback systems can evolve into black~widow systems.
We apply a simple geometric argument (beaming) for the evaporation efficiency of companion stars and demonstrate that it can explain 
the strongly bimodal distribution of the observed systems quite well.  

The outline of the paper is as follows: In Section 2, we introduce the details of our formation scenario. In Section 3, we compute the evolution of binary systems 
in the framework of this scenario. A discussion and a brief summary are given in Section 4.

\section{Formation scenario and binary evolution code}\label{sec:scenario}
Several ingredients must be considered for the formation of an eclipsing MSP system:
The NS must be able to spin~up to a millisecond period and the pulsar must turn on as a radio MSP;
the companion star should be ablated in order to produce orbital eclipses, and the orbital period and the companion star mass 
must match the observed values.
We find that these ingredients can be satisfied by the evolution of CV-like LMXBs. Similar to the evolution of CVs, 
the companion star becomes fully convective when its mass decreases to $0.2-0.3\,M_{\odot}$ \citep[e.g.][]{rvj83,ps89,prp02}. 
As a result, we assume that the magnetic braking stops operating and hence the mass transfer is temporarily halted \citep{rvj83,kin88}. 
At this time, the NS has accreted about $0.35\,M_{\odot}$ (assuming that the mass accumulation efficiency is 0.5 and the typical initial donor mass is $1\,M_{\odot}$), 
which is more than enough to spin up the NS to become an MSP \citep{tlk12}. 
When the NS turns on as a radio MSP (once the expanding magnetospheric boundary crosses the light-cylinder radius during termination of the mass transfer) 
it begins to evaporate its companion star \citep{vv88,rst89}
 by its pulsar wind of TeV ($e^-,e^+$) particles -- either directly or indirectly via secondary synchrotron MeV $\gamma$-rays produced near
the companion star. Even if the companion star refills its Roche~lobe, as a result of adiabatic expansion due to mass loss by evaporation, the
radio ejection mechanism \citep{krst88,bpd+01} will prevent accretion onto the MSP    
and expel the mass transferred from the companion star. This mass loss, if strong enough, may account for the orbital eclipse 
of the radio signal by free-free absorption.

In order to test this formation scenario, we have carried out detailed binary evolution calculations with a newly developed
Henyey evolutionary code, Modules for Experiments in Stellar Astrophysics ("MESA", see \citet{pbd+11,pcab+13} for details). 
In the binary evolution calculations, we included orbital angular momentum loss due to mass loss, gravitational wave radiation and magnetic braking. 
Previous studies indicated that a circumbinary (CB) disk can effectively extract angular momentum \citep{st01} and greatly affect 
the evolution of the binary system \citep{cl06}. 
Hence, we added this possibility and also included the effect of irradiation-induced mass loss from the companion star (as explained below).

More specifically, we assumed that the accretion efficiency of the NS is 0.5 and the mass lost from the system takes the specific angular momentum 
of the NS before the radio emission turns on \citep{prp02}. After the radio emission turns on, no mass will
be accreted by the NS because of the dominating pulsar radiation pressure, causing the transferred material to be expelled from the neighbourhood of the 
inner Lagrangian point \citep[e.g.][]{bpd+01}. Regarding the angular momentum loss due to gravitational wave radiation, we adopt the formula given by \citep{ll71}: 
\begin{equation}
\frac{dJ_{\rm GR}}{dt} = -\frac{32}{5}\frac{G^{7/2}}{c^{5}}\frac{M_{\rm NS}^{2}M_{2}^{2}(M_{\rm NS}+M_{2})^{1/2}}{a^{7/2}},
\end{equation}
where $M_{\rm NS}$ and $M_2$ denoted the NS and the companion star mass, $a$ is the semi-major axis of the (circular) orbit,
$c$ is the speed of light in vacuum and $G$ is the gravitational constant. 

To compute the angular momentum loss due to magnetic braking, we adopt the prescription of \citet{rvj83} [eq.~(36) with $\gamma = 4$]:
\begin{equation}
\frac{dJ_{\rm MB}}{dt} = -3.8 \times 10^{-30}M_{2}\,R_2^{4}\,\omega^{3} \quad{\rm dyn}\;{\rm cm},
\end{equation}
where $R_2$ is the radius of the donor star and $\omega$ is equal to the orbital angular velocity of the binary system.

During the evolution, we assume that a small part of mass lost from the companion is injected into a CB disk. 
The rate of CB disk mass injection is $\dot{M}_{\rm CB} = -\delta\dot{M}_{2}$. Assuming that the disk follows Keplerian rotation, the angular velocity of the 
disk is less than that of the binary system. Hence the CB disk will exert tidal torques on the binary and extract angular momentum from the system. 
To calculate the angular momentum loss due to a CB disk, we adopt the prescription proposed by \citet{st01} [see also eq.~(28) in \citet{sl12}]:
\begin{equation}
\dot{J}_{\rm CB} = -\gamma(\frac{2\pi a^{2}}{P})\dot{M}_{\rm CB}(\frac{t}{t_{\rm vi}})^{1/3},
\end{equation}
where $t$ is the time since mass transfer begins, $\gamma^{2}=r_{\rm i}/a$, $t_{\rm vi}=\frac{2\gamma^{3}P}{3\pi\alpha\beta^2}$ and where $r_{\rm i}$ is the inner radius of the disk, $\alpha$ is the viscosity parameter, $\beta=\frac{H_{\rm i}}{r_{\rm i}}$ and $H_{i}$ is the scaleheight of the disc \citep{ss73}. In the following calculation, we set $\gamma^{2}=1.7$ \citep{mm06}, $\alpha=0.01$, $\beta = 0.03$ \citep{bshh+04} and $\delta = 0.005$.

For the evaporation wind driven by the pulsar irradiation, a simple prescription was proposed by \citep{srp92}:
\begin{equation}
\dot{M}_{2,\rm evap}=-\frac{f}{2v_{2,\rm esc}^2}L_{\rm P}(\frac{R_{2}}{a})^2,
\label{eq:srp}
\end{equation}
where the pulsar's spin-down luminosity is given by: $L_{\rm p}=4\pi^2I\dot{P}/P^{3}$ ($I$ is the pulsar moment of inertia, $P$
is the spin period of the NS, and $\dot{P}$ is its spin down rate), 
 $v_{\rm 2,esc}$ is the escape velocity of a thermal wind from the surface of the companion star, and $f$, whose value is not well determined, 
is an efficiency factor. 
However, the implicit assumption of an isotropic energy flux leaving the pulsar is not valid if the outflow of the relativistic pulsar ($e^-,e^+$) wind
follows the geometry of the magnetosphere (and if this wind constitutes a significant part of the spin-down torque acting on the pulsar compared to
the magnetodipole radiation which is approximately emitted without much directivity). Thus we propose that there is a beaming effect which depends on the orientation 
of the pulsar B-field axis with respect to the direction of the companion star. This leads to a variety of evaporation efficiencies in systems which
may otherwise have similar characteristics with respect to $M_2, P_{\rm orb}, P$ and $\dot{P}$.
To simplify the description, we can still adopt Eq.(\ref{eq:srp}) under the assumption that $f$ is now a (free) parameter which takes {\it different}
values for different binaries and which depends on the geometry of the system -- i.e. a "beaming efficiency parameter" which may in principle 
reach a value above unity in the working frame of Eq.(\ref{eq:srp}).\\   
As we shall demonstrate, the value of $f$ is rather important for the
outcome of our calculations. For a recent discussion of the fraction of the incident spin-down luminosity being used for reprocessing 
to increase the dayside temperature of companion stars in eclipsing MSPs, see \citet{bvr+13}.

To calculate $L_{\rm p}$ we assumed for all evolutionary tracks an initial spin period, $P_0=3\;{\rm ms}$ and an initial 
period derivative, $\dot{P}_0=1.0\times 10^{-20}\;{\rm s}\,{\rm s}^{-1}$ (corresponding roughly to a surface magnetic filed strength of $\sim 10^{8}\;G$) 
at the epoch of radio MSP turn-on. $L_{\rm p}$ was then calculated as a function of time by assuming evolution with a constant braking index $n=3$, 
and using $I=10^{45}\;{\rm g}\,{\rm cm}^2$.

\section{Results of binary evolution calculations}
\begin{figure}
   \plotone{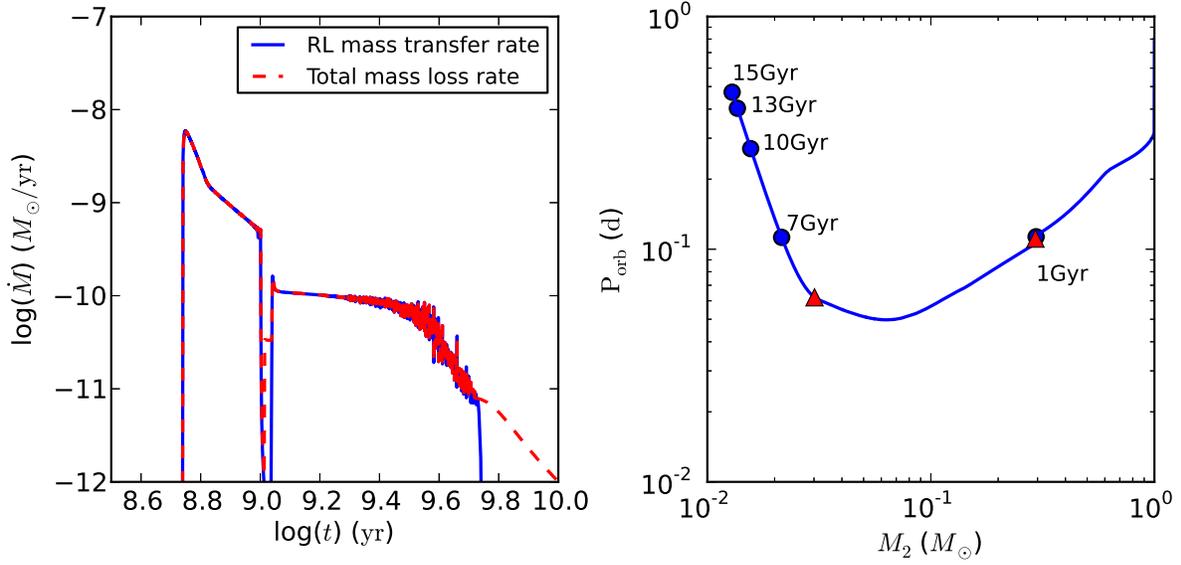}
   \caption{Evolution of a binary system with $M_{\rm NS}=1.4M_{\odot}$, $M_2=1.0M_{\odot}$, initial orbital period $P_{\rm orb}=0.8\;{\rm days}$ 
            and $f=0.02$. Left: evolution of the mass-transfer rate (blue line) and the total mass-loss rate (red, dashed line). 
            Right: evolution of the donor mass and the orbital period. The age of the system is indicated at certain epochs. 
            Detachments from the Roche lobe are marked by a triangle -- see text.}

\end{figure}

\begin{figure}
   \plotone{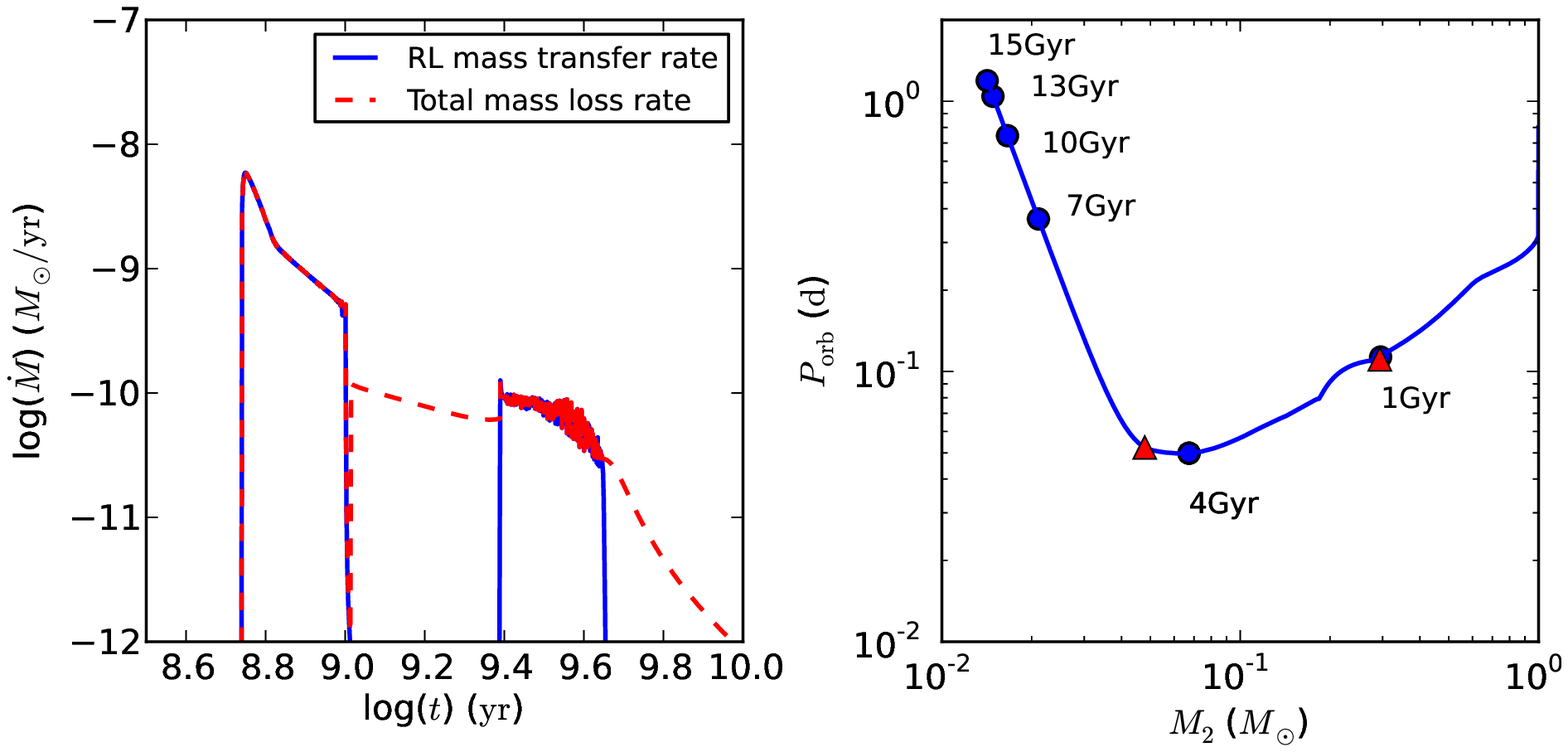}
   \caption{Similar to Figure~1 but for $f=0.07$ }
\end{figure}

\begin{figure}
   \plotone{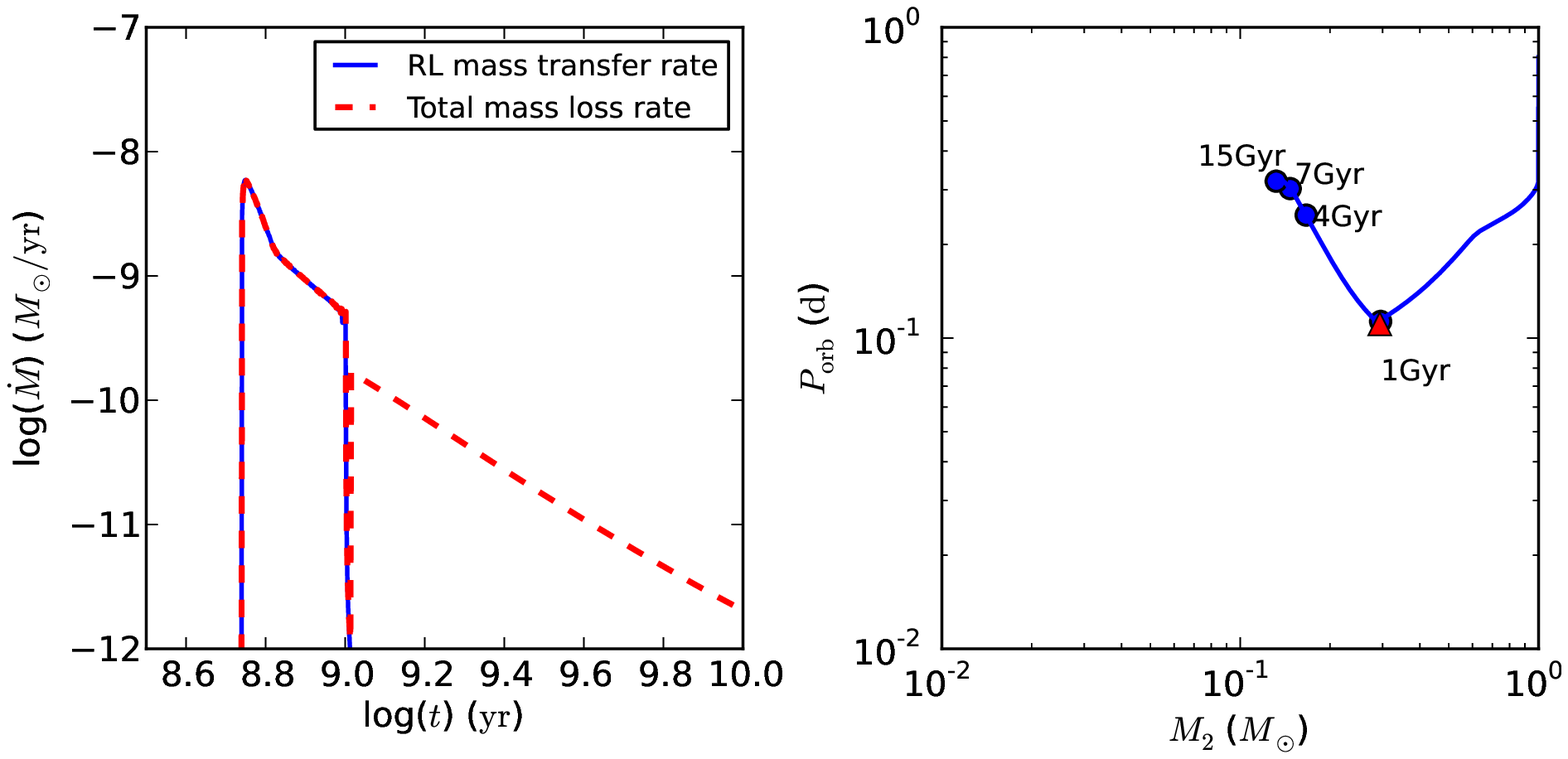}
   \caption{Similar to Figure~1 but for $f=0.10$}
\end{figure}

\begin{figure}
\plotone{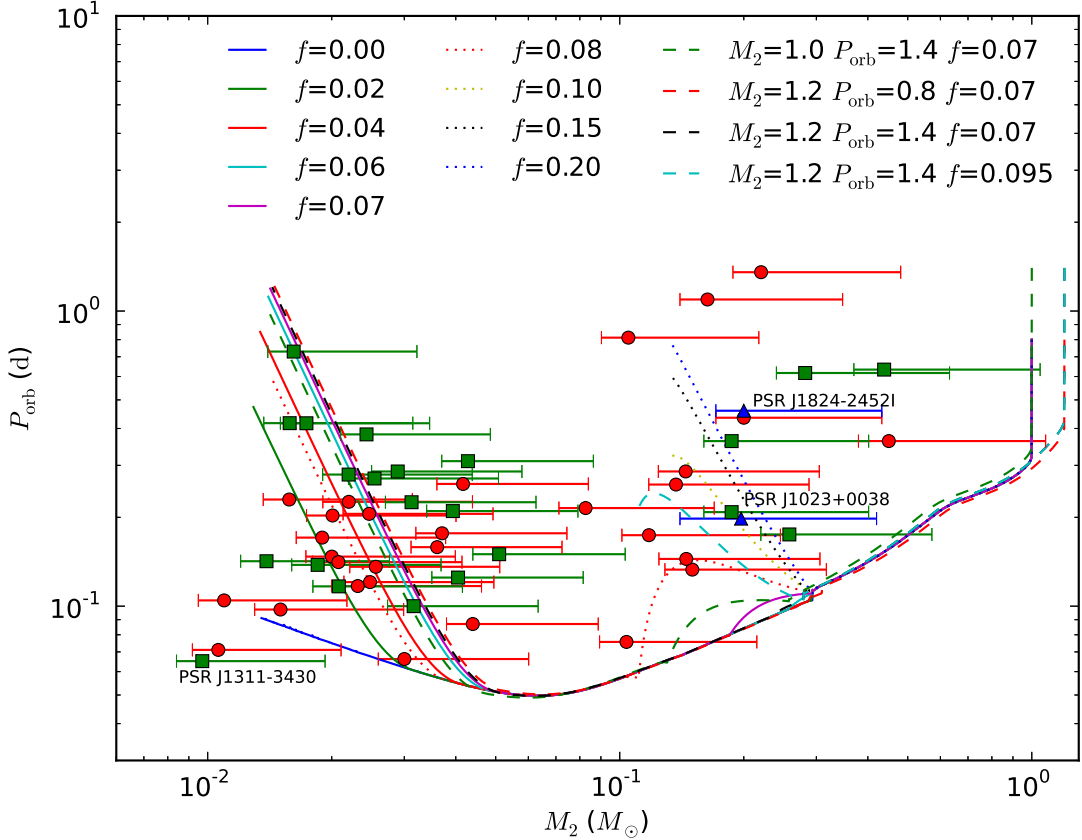}
\caption{Evolution of various binary systems calculated with different values of $f$. The observed data shows the 
black~widows ($M_2 \ll 0.1\;M_{\odot}$) and the redbacks ($M_2 \simeq 0.1-0.4\;M_{\odot}$). For the error~bars, 
the left and right ends correspond to an orbital inclination angle of $90\,^{\circ}$ and $25.8\,^{\circ}$ (the 90\% probability limit), respectively, for $M_{\rm NS} = 1.40 M_{\odot}$. 
The data for Galactic field (green squares) and globular cluster sources (red circles) was 
taken from the {\it ATNF Pulsar Catalogue} \texttt{http://www.atnf.csiro.au/research/pulsar/psrcat} \citep{mhth05}, 
P.~Freire's webpage \texttt{http://www.naic.edu/$\sim$pfreire/GCpsr.html}, \citet{rob13}, and references therein. 
PSR~J1023+0038 and J1824$-$2452I are marked with a blue triangle.}
\end{figure}

\begin{figure}
  \plotone{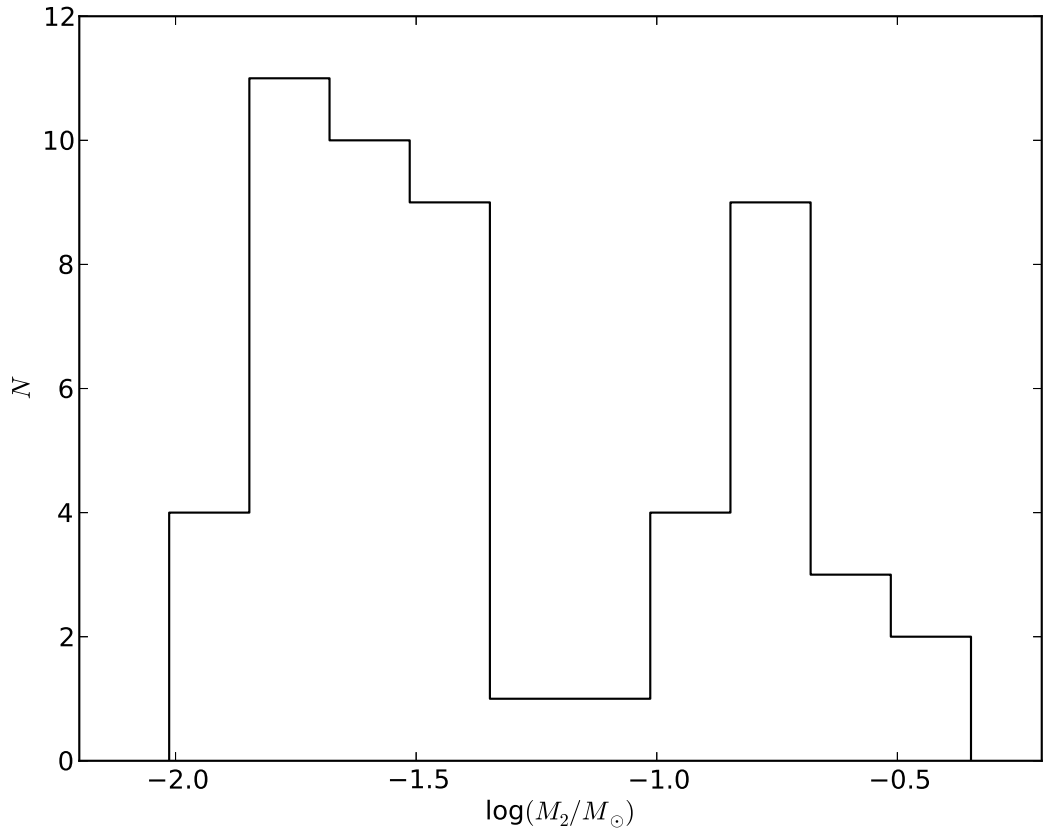}
   \caption{Histogram of the mass distribution of eclipsing MSPs. The bimodal splitting into the populations of black~widows (left) and
            redbacks (right) is clearly seen. According to \citep{frei05}, all black-widow-like systems may be evaporating systems, but in some cases low orbital inclinations prevent detection of eclipses}
\end{figure}

Figures~1--3 show the evolution of a binary system consisting of a $1.4M_{\odot}$ NS, a $1.0M_{\odot}$ donor star with an initial
orbital period of $0.8\;{\rm days}$ and for 
$f=0.02$, $f=0.07$, $f=0.10$, respectively. We end the calculations when the age exceeds 15~Gyr. 
The donor star begins mass transfer while it is still on the main sequence ($P_{\rm orb}\simeq 0.3\;{\rm days}$). 
Because of angular momentum loss, mainly due to magnetic braking (initially gravitational wave radiation is less important), 
the orbital period continues to decrease as the donor mass decreases.  
When the donor mass decreases to about $0.3M_{\odot}$, and the orbital period has decreased to about $2.5\;{\rm hr}$, 
we find that the donor star becomes fully convective, leading to Roche~lobe decoupling and switch-on of the radio pulsar and subsequent evaporation
of the companion star (see Section~\ref{sec:scenario}).

From this point onwards, the evolution of the binary system deviates with different values of $f$. This is a result of the competition between the 
widening of the orbit due to expelled, evaporated material and the shrinking
of the orbit due to gravitational wave radiation.
In Figure~1, the donor star will refill its Roche~lobe and resume mass transfer again after 57~Myr. 
However, as mentioned previously, the radio ejection mechanism will prevent the
NS from accreting at this stage.
In Figure~2, the donor star is decoupled from its Roche~lobe for a long time as a consequence of using a higher $f$-value. The more efficient evaporation of
the companion star causes the system to shrink less fast and the system remains detached for 1.4~Gyr.
Finally, in Figure~3, the donor star remains detached from its Roche~lobe at all times after the pulsar emission turns on. 
For different LMXBs with different initial parameters ($M_2, M_{\rm NS}, P_{\rm orb}$), 
we find that the shape of the evolution tracks are rather similar. 

In order to compare with observations, we calculated the evolution of several binary systems with different 
values of $f$. Figure~4 shows the evolution of constant initial donor star mass ($1.0\;M_{\odot}$) and orbital period ($0.80\;{\rm days}$) but 
with different values of $f$ between 0.0 and 0.20. Also plotted are four examples of binary systems with 1.0 and $1.2\;M_{\odot}$ donor stars 
and initial orbital periods of 0.8 and 1.4~days. (In addition, we calculated tracks for a $1.6\;M_{\odot}$ donor star
with practically the same outcome and thus not plotted here.) 
From this figure, it is clear that we can explain the formation of most of the eclipsing binary pulsars. 
In addition, we find that redbacks are more easily produced with larger values of $f$, while black~widows require smaller values of $f$. 
This means that a larger fraction of the pulsar luminosity is absorbed by the companion star in redback systems compared with the companions in black~widow systems. 
There is apparently a bifurcation in the evolutionary tracks at about $f\simeq 0.08$ ($f\simeq 0.10$ for $1.6\;M_{\odot}$ donor stars), 
such that systems with larger $f$-values produce redbacks and systems with smaller $f$-values lead to black~widows. 
Furthermore, we notice that the evolutionary tracks leading to the redbacks (the dotted tracks in Figure~4), in general, do not evolve into black~widow systems.
Only in a few cases with $f$-values fine-tuned near the bifurcation value ($f\simeq 0.08$), will redback systems with $P_{\rm orb}<4\;{\rm hr}$ evolve 
into black~widow systems.

\section{Discussion and Conclusion}
\subsection{Formation of redbacks vs. black~widows}
In Figure~5 we plot a histogram of the distribution of companion masses in eclipsing MSP systems. It is clearly seen that
the distribution is bimodal, with black~widow companions being lighter than redback companions (see also \citet{frei05} and \citet{rob13} 
for globular cluster and Galactic field sources, respectively).
We have demonstrated that this strongly bimodal distribution is caused by a bifurcation in the evolution of binaries depending on the
efficiency of companion evaporation following pulsar switch-on when the donor star detaches from its Roche-lobe.
We propose that the reason for the different values of $f$ may simply be related to the distribution of angles between the orbital angular momentum axis and the 
pulsar magnetic axis, which determines the geometry (beaming) of the outflow of charged particles from the pulsar magnetosphere.  Unfortunately, for MSPs it is very difficult to determine the magnetic inclination angle from polarization measurements and test this hypothesis.  
Observations of $\gamma$-ray detected MSPs show large variations in the conversion of pulsar spin-down luminosity
into $\gamma$-rays \citep[e.g.][]{rrc+11} which could also affect the evaporation efficiency. 

As seen in Figure~4, a few of the redback systems with relatively large companion masses and large orbital periods are not reached by our 
evolutionary tracks. We suggest that these systems somehow became (temporarily) detached when $M_2\sim\! 0.5\;M_{\odot}$, 
resulting in radio pulsar switch-on and ablation of the donor star at an early stage. Whether this is the result of a decrease 
in the effect of magnetic braking or a temporary radial contraction of the donor star, for example, due to a chemical abundance 
discontinuity in the hydrogen burning layer, is unknown.
Another possibility is accretion-powered radiation of the companion star. It has been argued that MeV $\gamma$-rays may be produced
in the inner region of the accretion disk and that such high-energy photons deposit more energy
in the outer layers of the irradiated stellar atmosphere than any other kind of illumination \citep{krst88,rste89}.
According to \cite{ham96}, however, the secular evolution of LMXBs is not much different from that of unilluminated systems
(see also brief discussion in Section~\ref{subsec:isolatedMSPs}).

\subsection{Comparison with observations}
In some cases, the spectra of eclipsing MSPs display prominent H$\alpha$ emission lines \citep{kh88,kbv+13}. 
In addition, some optical studies show that Roche-lobe filling factors are $0.4 \sim 1.0$ and that the companions are being ablated \citep{svlk99,vbk11,bvr+13}. 
These evidences show that the companions are probably non-degenerate stars instead of white dwarfs. In our models, the companions are indeed often  
hydrogen-rich non- or semi-degenerate stars, consistent with the observations. For the black~widow systems, the donors may, in principle, be hydrogen deficient.

PSR~J1023+0038 is a redback system, which is regarded as a "missing link" of the recycling formation scenario, discovered in a Green Bank Telescope drift 
scan pulsar survey by \citet{asr+09}. This binary MSP was revealed to be spinning at a period of 1.69~ms while orbiting a $\sim 0.2\;M_{\odot}$ companion star
with an orbital period of 4.8~hr.
In 2000--2001, the optical spectra showed double peaked emission lines indicating the 
presence of an accretion disk \citep{wat+09}\footnote{However, only absorption lines were observed in 2004, showing that the accretion disk had vanished.}. 
This indicates that the companion was previously overflowing its Roche~lobe and transferring mass, forming 
an accretion disk around the NS. 
However, there was no strong evidence of accretion-powered X-ray emission, possibly as a result
of a propeller effect during this epoch \citep{tau12}. 
The eclipsing nature of PSR~J1023+0038, its orbital period of 4.8~hr and a $0.14-0.42\;M_{\odot}$ companion star are all consistent with
the evolutionary tracks presented here (cf. Figure~3, and the dotted tracks in Figure~4).
It is very likely that this source is currently undergoing
cyclic episodes of accretion, possibly related to a combination of donor star irradiation \citep{br04} and accretion disk instabilities \citep{dlhc99}
combined with the radio ejection mechanism (L.~Burderi, private communication). The probability of catching
this system in the act of a one time transition from an LMXB to a radio MSP is simply too small.  

Indeed, similar evidence is found in another redback system: PSR~J1824$-$2452I in M28 \citep{beg06}. 
This source has recently been discovered to undergo cyclic transitions between accretion- and rotation powered states \citep{pfe+13}.
The MSP has an orbital period of 11.0~hr and a median (minimum) companion mass of $M_2 \simeq 0.20\;M_{\odot}$ ($0.17\;M_{\odot}$) for an assumed
NS mass of $1.35\;M_{\odot}$.
According to the ($M_{\rm WD},P_{\rm orb}$)--relation for helium white dwarfs formed in LMXBs \citep[e.g.][]{ts99} this companion star may
still have a few $0.01\;M_{\odot}$ of hydrogen left which forms the bloated envelope around the degenerate helium core.
Whether or not the system will continue evolving on a long time-scale of several Gyr with a bloated donor (cf. Figure~3) or finally detach and 
leave behind a helium white dwarf is unclear. In the latter case, the secular evolution (Myr) may lead to a spin period increase of this MSP 
depending on the timescale at which the magnetosphere expands due to the decreasing mass-transfer rate \citep{tau12}. 
However, on much shorter timescales (years or decades) several small torque reversals
arise from alternating phases of accretion and radio ejection which will cause a rather erratic behaviour of the spin period evolution\footnote{The reason for
the magnetospheric boundary being pushed back and forth with respect to the light cylinder may be related to thermal viscous instabilities in the 
accretion disk \citep{dlhc99}, disk-magnetosphere instabilities \citep{st93,nbc+97}, instabilities related to the irradiation of the donor \citep{br04}, 
a warped disk or other changes in the disk geometry/flow \citep{vcpw98,ywv97}, or possibly clumps in the 
material transfered from the convective envelope of the companion.}.

PSR~J1816+4510 (K.~Stovall et~al., in preparation) is yet another interesting example illustrating the puzzling nature of a redback system. 
\citet{kbv+13} recently reported optical spectroscopy of this system and demonstrated that, on the one hand the companion star
possesses the characteristic features of a helium white dwarf spectrum while on the other hand, 
it is metal-rich and has a low surface gravity 
-- typical to that of a much larger, bloated non- or semi-degenerate star (causing the ionized gas eclipses of the radio pulsar signal).

PSR~J1719$-$1438 \citep{bbb+11} has an intriguing formation history. If the very low-mass companion (planet) has a composition of helium, 
then the system could have formed via the converging LMXB scenario as described here \citep[as demonstrated in detail by ][]{bdh12}. 
However, if the composition is made of carbon and oxygen
then the formation of this systems is most likely the result of an ultra-compact X-ray binary \citep[UCXB,][]{vnvj12b}; a scenario in which
an intermediate-mass carbon-oxygen white dwarf transfers, basically, all its mass towards a NS on a Hubble timescale, while the system widens. 
The eclipsing black~widow PSR~J1311$-$3430, recently discovered by \citet{pgf+12} and \citet{rfs+12}, has $P_{\rm orb}=93\;{\rm min}$ and a minimum mean density
of $45\;{\rm g}\,{\rm cm}^{-3}$, which hints that this system may also have evolved from an UCXB. This may explain its location slightly
below our evolutionary tracks in Figure~4. For a study on the formation of black~widow-like MSPs in globular clusters from UCXBs, see also \citet{rpr00}.

In general, it is not straight forward to compare our estimated mass-loss rates from modelling 
(and the predicted rates of orbital widening) with those inferred from observations.
In particular, in the classical black~widow systems PSR~1957+20 and J2051$-$0827 where there is observational evidence \citep{aft94,lvt+11} 
for spin-orbit couplings and tidal dissipation leading to changes in the gravitational quadrupole moment of the companion \citep{as94}.
These effects result in severe orbital evolution, even with sign changes of the orbital period derivative, thereby making a comparison
to our long-term evolution models difficult.

\subsection{Formation of isolated MSPs} \label{subsec:isolatedMSPs}
At present, almost 300 MSPs (here defined with $P<30\;{\rm ms}$) are detected in the Galactic field (154) and in globular clusters (130). 
These MSPs are thought to be formed by the recycling process. 
However, about 1/3 of all MSPs are isolated. It is still not well known how these solitary MSPs are formed. 
It was suggested by \citet{vv88} that isolated MSPs are formed from black~widows which ablate away their companions by the 
energetic particles and the $\gamma$-rays emanating from radio MSPs. 
Furthermore, there is evidence for such a scenario by the discoveries of MSPs with planets \citep{wf92}, 
and even indications of an MSP with an asteroid 
belt \citep{scm+13}, possibly the debris of a former tidally disrupted planet-like companion.

From Figures~1 and 2, it seems that it is difficult to produce isolated MSPs within a Hubble time 
from our calculations of the converging LMXB scenario. 
However, we have not considered the structure change of the companion star due to the penetration of radiation into its envelope. 
As \citet{pod91} suggested, the radius of such a companion will greatly expand. Hence, the mass-loss rate due to an evaporation wind could increase 
significantly and the evaporation timescale of the companion should be shorter. \citet{bdh12} did include irradiation feedback of the donor star
but find that this effect (while causing cyclic mass-transfer episodes) does not induce any 
remarkable changes in the evolution in the ($M_2,P_{\rm orb}$)--plane. 
To judge from their figure~1, the evolution is accelerated slightly with respect to time 
and it may still be possible for the pulsar to ablate away the companion and produce an isolated MSP.

To conclude, we have demonstrated that the two populations of eclipsing binary MSPs, the black~widows and the redbacks,
are not only distinct with respect to their observational properties, but also from an evolutionary point of view. 
The determining factor for producing black~widows or redbacks is the efficiency of the companion evaporation. Our calculations suggest
that the fraction of pulsar spin-down luminosity absorbed by the companion is larger for redbacks than that for black~widows
(possibly as a result of a different beaming geometry and/or MSP $\gamma$-ray luminosity).
Despite our rather simple modelling, we are able to reproduce the distribution of the observed systems quite well. 
Last but not least, we find that redback systems do not evolve into black~widow systems with time.

\begin{acknowledgements}
We are grateful to Norbert Langer for helpful discussions. We thank the anonymous referee for suggestions that improved our paper, 
and also Philipp Podsiadlowskii for discussions on the evaporation efficiency parameter.
This work was partly supported by NSFC (Nos. 10973036, 11173055, 11033008 and 11003003), the CAS (No. KJCX2-YW-T24, the Talent Project of Western Light) 
and the Talent Project of Young Researchers of Yunnan Province (No. 2012HB037). 
 
\end{acknowledgements}

\bibliographystyle{apj}

\begin{thebibliography}{67}
\expandafter\ifx\csname natexlab\endcsname\relax\def\natexlab#1{#1}\fi

\bibitem[{{Abdo} {et~al.}(2009){Abdo}, {Ackermann}, {Ajello}, {Atwood}, \&
  {et~al.}}]{aaa+09}
{Abdo}, A.~A., {Ackermann}, M., {Ajello}, M., {Atwood}, W.~B., \& {et~al.}
  2009, Science, 325, 848, 848

\bibitem[{Alpar {et~al.}(1982)Alpar, Cheng, Ruderman, \& Shaham}]{acrs82}
Alpar, M.~A., Cheng, A.~F., Ruderman, M.~A., \& Shaham, J. 1982, \nat, 300,
  728, 728

\bibitem[{{Applegate} \& {Shaham}(1994)}]{as94}
{Applegate}, J.~H., \& {Shaham}, J. 1994, \apj, 436, 312, 312

\bibitem[{{Archibald} {et~al.}(2009){Archibald}, Stairs, Ransom, Kaspi,
  Kondratiev, Lorimer, McLaughlin, Boyles, Hessels, Lynch, van Leeuwen,
  Roberts, Jenet, Champion, Rosen, Barlow, Dunlap, \& Remillard}]{asr+09}
{Archibald}, A.~M., Stairs, I.~H., Ransom, S.~M., {et~al.} 2009, Science, 324,
  1411, 1411

\bibitem[{{Arzoumanian} {et~al.}(1994){Arzoumanian}, {Fruchter}, \&
  {Taylor}}]{aft94}
{Arzoumanian}, Z., {Fruchter}, A.~S., \& {Taylor}, J.~H. 1994, \apjl, 426, L85,
  L85

\bibitem[{{Bailes} {et~al.}(2011){Bailes}, {Bates}, {Bhalerao}, {Bhat},
  {Burgay}, {Burke-Spolaor}, {D'Amico}, {Johnston}, {Keith}, {Kramer},
  {Kulkarni}, {Levin}, {Lyne}, {Milia}, {Possenti}, {Spitler}, {Stappers}, \&
  {van Straten}}]{bbb+11}
{Bailes}, M., {Bates}, S.~D., {Bhalerao}, V., {et~al.} 2011, Science, 333,
  1717, 1717

\bibitem[{{B\'egin}(2006)}]{beg06}
{B\'egin}, S. 2006, Master's thesis, Faculty of Physics, Univ. British
  Columbia,

\bibitem[{{Belle} {et~al.}(2004){Belle}, {Sanghi}, {Howell}, {Holberg}, \&
  {Williams}}]{bshh+04}
{Belle}, K.~E., {Sanghi}, N., {Howell}, S.~B., {Holberg}, J.~B., \& {Williams},
  P.~T. 2004, \aj, 128, 448, 448

\bibitem[{{Benvenuto} {et~al.}(2012){Benvenuto}, {De Vito}, \&
  {Horvath}}]{bdh12}
{Benvenuto}, O.~G., {De Vito}, M.~A., \& {Horvath}, J.~E. 2012, \apjl, 753,
  L33, L33

\bibitem[{Bhattacharya \& {van den Heuvel}(1991)}]{bv91}
Bhattacharya, D., \& {van den Heuvel}, E. P.~J. 1991, Physics Reports, 203, 1,
  1

\bibitem[{{Breton} {et~al.}(2013){Breton}, {van Kerkwijk}, {Roberts},
  {Hessels}, {Camilo}, {McLaughlin}, {Ransom}, {Ray}, \& {Stairs}}]{bvr+13}
{Breton}, R.~P., {van Kerkwijk}, M.~H., {Roberts}, M.~S.~E., {et~al.} 2013,
  ArXiv e-prints, arXiv:1302.1790

\bibitem[{{B{\"u}ning} \& {Ritter}(2004)}]{br04}
{B{\"u}ning}, A., \& {Ritter}, H. 2004, \aap, 423, 281, 281

\bibitem[{{Burderi} {et~al.}(2001){Burderi}, {Possenti}, {D'Antona}, {Di
  Salvo}, {Burgay}, {Stella}, {Menna}, {Iaria}, {Campana}, \&
  {d'Amico}}]{bpd+01}
{Burderi}, L., {Possenti}, A., {D'Antona}, F., {et~al.} 2001, \apjl, 560, L71,
  L71

\bibitem[{{Chen} \& {Li}(2006)}]{cl06}
{Chen}, W.-C., \& {Li}, X.-D. 2006, \mnras, 373, 305, 305

\bibitem[{{Cumming} {et~al.}(2001){Cumming}, {Zweibel}, \& {Bildsten}}]{czb01}
{Cumming}, A., {Zweibel}, E., \& {Bildsten}, L. 2001, \apj, 557, 958, 958

\bibitem[{{Dubus} {et~al.}(1999){Dubus}, {Lasota}, {Hameury}, \&
  {Charles}}]{dlhc99}
{Dubus}, G., {Lasota}, J.-P., {Hameury}, J.-M., \& {Charles}, P. 1999, \mnras,
  303, 139, 139

\bibitem[{{Ergma} \& {Fedorova}(1992)}]{ef92}
{Ergma}, E., \& {Fedorova}, A.~V. 1992, \aap, 265, 65, 65

\bibitem[{{Freire}(2005)}]{frei05}
{Freire}, P.~C.~C. 2005, in Astronomical Society of the Pacific Conference
  Series, Vol. 328, Binary Radio Pulsars, ed. F.~A. {Rasio} \& I.~H. {Stairs},
  405

\bibitem[{{Fruchter} {et~al.}(1988){Fruchter}, {Stinebring}, \&
  {Taylor}}]{fst88}
{Fruchter}, A.~S., {Stinebring}, D.~R., \& {Taylor}, J.~H. 1988, \nat, 333,
  237, 237

\bibitem[{{Hameury}(1996)}]{ham96}
{Hameury}, J.-M. 1996, \aap, 305, 468, 468

\bibitem[{{Kaplan} {et~al.}(2013){Kaplan}, {Bhalerao}, {van Kerkwijk},
  {Koester}, {Kulkarni}, \& {Stovall}}]{kbv+13}
{Kaplan}, D.~L., {Bhalerao}, V.~B., {van Kerkwijk}, M.~H., {et~al.} 2013, \apj,
  765, 158, 158

\bibitem[{{King}(1988)}]{kin88}
{King}, A.~R. 1988, \qjras, 29, 1, 1

\bibitem[{{King} {et~al.}(2005){King}, {Beer}, {Rolfe}, {Schenker}, \&
  {Skipp}}]{kbr+05}
{King}, A.~R., {Beer}, M.~E., {Rolfe}, D.~J., {Schenker}, K., \& {Skipp}, J.~M.
  2005, \mnras, 358, 1501, 1501

\bibitem[{{King} {et~al.}(2003){King}, {Davies}, \& {Beer}}]{kdb03}
{King}, A.~R., {Davies}, M.~B., \& {Beer}, M.~E. 2003, \mnras, 345, 678, 678

\bibitem[{{Kluzniak} {et~al.}(1988){Kluzniak}, {Ruderman}, {Shaham}, \&
  {Tavani}}]{krst88}
{Kluzniak}, W., {Ruderman}, M., {Shaham}, J., \& {Tavani}, M. 1988, \nat, 334,
  225, 225

\bibitem[{{Konar} \& {Bhattacharya}(1997)}]{kb97}
{Konar}, S., \& {Bhattacharya}, D. 1997, \mnras, 284, 311, 311

\bibitem[{{Kulkarni} \& {Hester}(1988)}]{kh88}
{Kulkarni}, S.~R., \& {Hester}, J.~J. 1988, \nat, 335, 801, 801

\bibitem[{{Landau} \& {Lifshitz}(1971)}]{ll71}
{Landau}, L.~D., \& {Lifshitz}, E.~M. 1971,

\bibitem[{{Lazaridis} {et~al.}(2011){Lazaridis}, {Verbiest}, {Tauris},
  {Stappers}, {Kramer}, {Wex}, {Jessner}, {Cognard}, {Desvignes}, {Janssen},
  {Purver}, {Theureau}, {Bassa}, \& {Smits}}]{lvt+11}
{Lazaridis}, K., {Verbiest}, J.~P.~W., {Tauris}, T.~M., {et~al.} 2011, \mnras,
  584, 584

\bibitem[{{Manchester} {et~al.}(2005){Manchester}, {Hobbs}, {Teoh}, \&
  {Hobbs}}]{mhth05}
{Manchester}, R.~N., {Hobbs}, G.~B., {Teoh}, A., \& {Hobbs}, M. 2005, \aj, 129,
  1993, 1993

\bibitem[{{Muno} \& {Mauerhan}(2006)}]{mm06}
{Muno}, M.~P., \& {Mauerhan}, J. 2006, \apjl, 648, L135, L135

\bibitem[{{Nelson} {et~al.}(1997){Nelson}, {Bildsten}, {Chakrabarty}, {Finger},
  {Koh}, {Prince}, {Rubin}, {Scott}, {Vaughan}, \& {Wilson}}]{nbc+97}
{Nelson}, R.~W., {Bildsten}, L., {Chakrabarty}, D., {et~al.} 1997, \apjl, 488,
  L117, L117

\bibitem[{{Pallanca} {et~al.}(2012){Pallanca}, {Mignani}, {Dalessandro},
  {Ferraro}, {Lanzoni}, {Possenti}, {Burgay}, \& {Sabbi}}]{pmd+12}
{Pallanca}, C., {Mignani}, R.~P., {Dalessandro}, E., {et~al.} 2012, \apj, 755,
  180, 180

\bibitem[{{Papitto} {et~al.}(2013){Papitto}, {Ferrigno}, {Bozzo}, {Rea},
  {Pavan}, {Campana}, {Romano}, {Burderi}, {Di Salvo}, {Riggio}, {Torres},
  {Falanga}, {Hessels}, {Burgay}, {Sarkissian}, {Wieringa}, {Filipovi{\'c}}, \&
  {Wong}}]{pfe+13}
{Papitto}, A., {Ferrigno}, C., {Bozzo}, E., {et~al.} 2013, astro-ph: 1305.3884

\bibitem[{{Paxton} {et~al.}(2011){Paxton}, {Bildsten}, {Dotter}, {Herwig},
  {Lesaffre}, \& {Timmes}}]{pbd+11}
{Paxton}, B., {Bildsten}, L., {Dotter}, A., {et~al.} 2011, \apjs, 192, 3, 3

\bibitem[{{Paxton} {et~al.}(2013){Paxton}, {Cantiello}, {Arras}, {Bildsten},
  {Brown}, {Dotter}, {Mankovich}, {Montgomery}, {Stello}, {Timmes}, \&
  {Townsend}}]{pcab+13}
{Paxton}, B., {Cantiello}, M., {Arras}, P., {et~al.} 2013, ArXiv e-prints,
  arXiv:1301.0319

\bibitem[{{Pletsch} {et~al.}(2012){Pletsch}, {Guillemot}, {Fehrmann}, \& {et
  al.}}]{pgf+12}
{Pletsch}, H.~J., {Guillemot}, L., {Fehrmann}, H., \& {et al.} 2012, Science,
  338, 1314, 1314

\bibitem[{{Podsiadlowski}(1991)}]{pod91}
{Podsiadlowski}, P. 1991, \nat, 350, 136, 136

\bibitem[{{Podsiadlowski} {et~al.}(2002){Podsiadlowski}, {Rappaport}, \&
  {Pfahl}}]{prp02}
{Podsiadlowski}, P., {Rappaport}, S., \& {Pfahl}, E.~D. 2002, \apj, 565, 1107,
  1107

\bibitem[{{Pylyser} \& {Savonije}(1989)}]{ps89}
{Pylyser}, E.~H.~P., \& {Savonije}, G.~J. 1989, \aap, 208, 52, 52

\bibitem[{{Ransom} {et~al.}(2011){Ransom}, {Ray}, {Camilo}, {Roberts}, {{\c
  C}elik}, {Wolff}, {Cheung}, {Kerr}, {Pennucci}, {DeCesar}, {Cognard}, {Lyne},
  {Stappers}, {Freire}, {Grove}, {Abdo}, {Desvignes}, {Donato}, {Ferrara},
  {Gehrels}, {Guillemot}, {Gwon}, {Harding}, {Johnston}, {Keith}, {Kramer},
  {Michelson}, {Parent}, {Saz Parkinson}, {Romani}, {Smith}, {Theureau},
  {Thompson}, {Weltevrede}, {Wood}, \& {Ziegler}}]{rrc+11}
{Ransom}, S.~M., {Ray}, P.~S., {Camilo}, F., {et~al.} 2011, \apjl, 727, L16,
  L16

\bibitem[{{Rappaport} {et~al.}(1983){Rappaport}, {Verbunt}, \& {Joss}}]{rvj83}
{Rappaport}, S., {Verbunt}, F., \& {Joss}, P.~C. 1983, \apj, 275, 713, 713

\bibitem[{{Rasio} {et~al.}(2000){Rasio}, {Pfahl}, \& {Rappaport}}]{rpr00}
{Rasio}, F.~A., {Pfahl}, E.~D., \& {Rappaport}, S. 2000, \apjl, 532, L47, L47

\bibitem[{{Roberts}(2013)}]{rob13}
{Roberts}, M.~S.~E. 2013, in IAU Symposium, Vol. 291, IAU Symposium, 127--132

\bibitem[{{Romani} {et~al.}(2012){Romani}, {Filippenko}, {Silverman}, {Cenko},
  {Greiner}, {Rau}, {Elliott}, \& {Pletsch}}]{rfs+12}
{Romani}, R.~W., {Filippenko}, A.~V., {Silverman}, J.~M., {et~al.} 2012, \apjl,
  760, L36, L36

\bibitem[{{Ruderman} {et~al.}(1989{\natexlab{a}}){Ruderman}, {Shaham}, \&
  {Tavani}}]{rst89}
{Ruderman}, M., {Shaham}, J., \& {Tavani}, M. 1989{\natexlab{a}}, \apj, 336,
  507, 507

\bibitem[{{Ruderman} {et~al.}(1989{\natexlab{b}}){Ruderman}, {Shaham},
  {Tavani}, \& {Eichler}}]{rste89}
{Ruderman}, M., {Shaham}, J., {Tavani}, M., \& {Eichler}, D.
  1989{\natexlab{b}}, \apj, 343, 292, 292

\bibitem[{{Shakura} \& {Sunyaev}(1973)}]{ss73}
{Shakura}, N.~I., \& {Sunyaev}, R.~A. 1973, \aap, 24, 337, 337

\bibitem[{{Shannon} {et~al.}(2013){Shannon}, {Cordes}, {Metcalfe}, {Lazio},
  {Cognard}, {Desvignes}, {Janssen}, {Jessner}, {Kramer}, {Lazaridis},
  {Purver}, {Stappers}, \& {Theureau}}]{scm+13}
{Shannon}, R.~M., {Cordes}, J.~M., {Metcalfe}, T.~S., {et~al.} 2013, ArXiv
  e-prints, arXiv:1301.6429

\bibitem[{{Shao} \& {Li}(2012)}]{sl12}
{Shao}, Y., \& {Li}, X.-D. 2012, \apj, 756, 85, 85

\bibitem[{{Spruit} \& {Taam}(1993)}]{st93}
{Spruit}, H.~C., \& {Taam}, R.~E. 1993, \apj, 402, 593, 593

\bibitem[{{Spruit} \& {Taam}(2001)}]{st01}
---. 2001, \apj, 548, 900, 900

\bibitem[{{Srinivasan} {et~al.}(1990){Srinivasan}, {Bhattacharya}, {Muslimov},
  \& {Tsygan}}]{sbmt90}
{Srinivasan}, G., {Bhattacharya}, D., {Muslimov}, A.~G., \& {Tsygan}, A.~J.
  1990, Current Science, 59, 31, 31

\bibitem[{{Stappers} {et~al.}(1999){Stappers}, {van Kerkwijk}, {Lane}, \&
  {Kulkarni}}]{svlk99}
{Stappers}, B.~W., {van Kerkwijk}, M.~H., {Lane}, B., \& {Kulkarni}, S.~R.
  1999, \apjl, 510, L45, L45

\bibitem[{{Stappers} {et~al.}(1996){Stappers}, {Bailes}, {Lyne}, {Manchester},
  {D'Amico}, {Tauris}, {Lorimer}, {Johnston}, \& {Sandhu}}]{sbl+96}
{Stappers}, B.~W., {Bailes}, M., {Lyne}, A.~G., {et~al.} 1996, \apjl, 465,
  L119, L119

\bibitem[{{Stevens} {et~al.}(1992){Stevens}, {Rees}, \&
  {Podsiadlowski}}]{srp92}
{Stevens}, I.~R., {Rees}, M.~J., \& {Podsiadlowski}, P. 1992, \mnras, 254, 19P,
  19P

\bibitem[{{Tauris}(2012)}]{tau12}
{Tauris}, T.~M. 2012, Science, 335, 561, 561

\bibitem[{{Tauris} {et~al.}(2012){Tauris}, {Langer}, \& {Kramer}}]{tlk12}
{Tauris}, T.~M., {Langer}, N., \& {Kramer}, M. 2012, \mnras, 425, 1601, 1601

\bibitem[{Tauris \& Savonije(1999)}]{ts99}
Tauris, T.~M., \& Savonije, G.~J. 1999, \aap, 350, 928, 928

\bibitem[{{van den Heuvel} \& {van Paradijs}(1988)}]{vv88}
{van den Heuvel}, E.~P.~J., \& {van Paradijs}, J. 1988, \nat, 334, 227, 227

\bibitem[{{van Haaften} {et~al.}(2012){van Haaften}, {Nelemans}, {Voss}, \&
  {Jonker}}]{vnvj12b}
{van Haaften}, L.~M., {Nelemans}, G., {Voss}, R., \& {Jonker}, P.~G. 2012,
  \aap, 541, A22, A22

\bibitem[{{van Kerkwijk} {et~al.}(2011){van Kerkwijk}, {Breton}, \&
  {Kulkarni}}]{vbk11}
{van Kerkwijk}, M.~H., {Breton}, R.~P., \& {Kulkarni}, S.~R. 2011, \apj, 728,
  95, 95

\bibitem[{{van Kerkwijk} {et~al.}(1998){van Kerkwijk}, {Chakrabarty},
  {Pringle}, \& {Wijers}}]{vcpw98}
{van Kerkwijk}, M.~H., {Chakrabarty}, D., {Pringle}, J.~E., \& {Wijers},
  R.~A.~M.~J. 1998, \apjl, 499, L27, L27

\bibitem[{{Wang} {et~al.}(2009){Wang}, {Archibald}, {Thorstensen}, {Kaspi},
  {Lorimer}, {Stairs}, \& {Ransom}}]{wat+09}
{Wang}, Z., {Archibald}, A.~M., {Thorstensen}, J.~R., {et~al.} 2009, \apj, 703,
  2017, 2017

\bibitem[{{Wolszczan} \& {Frail}(1992)}]{wf92}
{Wolszczan}, A., \& {Frail}, D.~A. 1992, \nat, 355, 145, 145

\bibitem[{{Yi} {et~al.}(1997){Yi}, {Wheeler}, \& {Vishniac}}]{ywv97}
{Yi}, I., {Wheeler}, J.~C., \& {Vishniac}, E.~T. 1997, \apjl, 481, L51, L51

\bibitem[{{Zhang}(1998)}]{zhan98}
{Zhang}, C.~M. 1998, \aap, 330, 195, 195

\end{thebibliography}


\end{document}